\documentclass[fleqn,usenatbib]{mnras}

\usepackage{newtxtext,newtxmath}

\usepackage[T1]{fontenc}
\usepackage{ae,aecompl}

\usepackage{graphicx}	
\usepackage{amsmath}	
\usepackage{bm}
\usepackage[usenames]{color}
\usepackage{animate}
\usepackage{mathrsfs}

\graphicspath{{Figures/}}
\usepackage[normalem]{ulem}
\usepackage{array}
\usepackage{framed}

\title[Cusp of cusps]{On the cusp of cusps: a universal model for extreme scattering events in the ISM}

\author[Jow et al.]{
Dylan L. Jow,$^{1,2,3}$\thanks{E-mail: djow@physics.utoronto.ca}
Ue-Li Pen,$^{4,1,2,3,5,6}$
and Daniel Baker$^{1,2}$
\\
$^{1}$Canadian Institute for Theoretical Astrophysics, University of Toronto, 60 St. George Street, Toronto, ON M5S 3H8, Canada\\
$^{2}$Department of Physics, University of Toronto, 60 St. George Street, Toronto, ON M5S 1A7, Canada\\
$^{3}$Dunlap Institute for Astronomy \& Astrophysics, University of Toronto, AB 120-50 St. George Street, Toronto, ON M5S 3H4, Canada\\
$^{4}$Institute of Astronomy and Astrophysics, Academia Sinica, Astronomy-Mathematics Building, No. 1, Section 4,
Roosevelt Road, Taipei 10617, Taiwan \\
$^{5}$Perimeter Institute for Theoretical Physics, 31 Caroline St. North, Waterloo, ON, Canada N2L 2Y5\\
$^{6}$Canadian Institute for Advanced Research, CIFAR program in Gravitation and Cosmology
}

\date{Accepted XXX. Received YYY; in original form ZZZ}

\pubyear{2023}

\begin{document}
\label{firstpage}
\pagerange{\pageref{firstpage}--\pageref{lastpage}}
\maketitle

\begin{abstract}
The scattering structures in the ISM responsible for so-called ``extreme scattering events" (ESEs), observed in quasars and pulsars, remain enigmatic. Current models struggle to explain the high-frequency light curves of ESEs, and a recent analysis of a double lensing event in PSR\,B0834+06 reveals features of ESEs that may also be challenging to accommodate via existing models. We propose that these features arise naturally when the lens has a cusp-like profile, described by the elementary $A_3$ cusp catastrophe. This is an extension of previous work describing pulsar scintillation as arising from $A_2$ fold catastrophes in thin, corrugated plasma sheets along the line of sight. We call this framework of describing the lens potentials via elementary catastrophes ``doubly catastrophic lensing", as catastrophes (e.g. folds and cusps) have long been used to describe universal features in the light curves of lensing events that generically manifest, regardless of the precise details of the lens. Here, we argue that the lenses themselves may be described by these same elementary structures. If correct, the doubly catastrophic lensing framework would provide a unified description of scintillation and ESEs, where the lenses responsible for these scattering phenomena are universal and can be fully described by a small number of unfolding parameters. This could enable their application as giant cosmic lenses for precision measurements of coherent sources, including FRBs and pulsars.
\end{abstract}




\section{Introduction}
\label{sec:intro}

The enigmatic extreme scattering events (ESEs) that were first discovered in quasars in the late 80s \citep{Fiedler1987} and in pulsars a few years later \citep{1993Natur.366..320C} have presented a long-standing mystery in observations of radio sources. While they are known to be caused by scattering in the interstellar medium (ISM), the precise form of the plasma structures that cause these events and the physical origin of these structures remains unknown. Interest in observing and understanding ESEs has increased, with recent work highlighting their relative ubiquity and setting the stage for future surveys of these mysterious events \citep{2016Sci...351..354B}. Moreover, the excess time delays induced by ESEs have implications for precision gravitational wave detection through pulsar timing arrays. Precise modelling of these excess delays will be necessary to move beyond detection of a stochastic gravitational wave background to individual detections \citep{2019A&ARv..27....5B}. Recently, novel phase retrieval techniques have been used for precision localization of the refractive images formed by the ESE lens \citep{2022arXiv220806884Z}. In conjunction with such techniques, new observations from current and next-generation radio telescopes built for pulsar timing arrays and fast radio burst (FRB) detections, among other purposes, will allow us to test the variety of models that have been proposed to explain ESEs.

While future observations will hopefully shed light on the plasma structures causing ESEs, current observations pose several theoretical challenges. Assuming ESEs are caused by three-dimensional plasma inhomogeneities in the ISM (i.e. an over- or under-dense cloud of ionized plasma) leads to inferences for the pressure of such clouds that are several orders of magnitude in excess of typical pressures in the diffuse ISM \citep{1998ApJ...496..253C}. If such highly pressurized clouds existed, then they would be unstable on the time-scales needed to explain ESE observations; this is known as the over-pressure problem. Thin, two-dimensional current sheets that are aligned with the line of sight have been proposed as a potential resolution to the over-pressure problem \citep{1987Natur.328..324R, 2012MNRAS.421L.132P}, but, thus far, such models struggle to explain certain features of the ESE light curves, in particular their rich frequency structure \citep{1998ApJ...498L.125W}. For example, while a two-dimensional Gaussian profile may fit the low-frequency light curves observed in ESEs, they fail to match the high-frequency light curves \citep{1998ApJ...496..253C}. Typically, such models invoke substructure in the ISM that becomes resolved at high frequencies to explain the complex morphologies of the high-frequency light curves; however, it remains desirable to be able to explain both the time and frequency structure of ESEs with a single lens model. Cold, self-gravitating clouds of neutral gas with an ionized skin have been proposed to explain the frequency structure of ESEs \citep{1995ApJ...441...70H, 1998ApJ...498L.125W}, but if correct would imply that a substantial fraction of the galaxy's mass is contained within these clouds. 

ESEs are not the only scattering phenomenon associated with the ISM that radio sources are observed to undergo. Pulsars are observed to scintillate due to multi-path scattering in the ISM. It has generally been assumed that pulsar scintillation and extreme scattering events are distinct phenomena, caused by different plasma structures in the ISM. However, just as thin plasma sheets have been proposed as an explanation for ESEs, in recent decades, there has been growing observational evidence that a substantial fraction of scintillation observations (if not all) can be explained by refractive plasma sheets along the line of sight \citep{2001ApJ...549L..97S, 2004MNRAS.354...43W, 2006ApJ...640L.159G, 2010ApJ...708..232B, 2014MNRAS.442.3338P}. This is in contrast to traditional models of an extended Kolmogorov turbulent medium. A similar story has been playing out in the study of the turbulent ISM through magnetohydrodynamic (MHD) simulations. Recent MHD simulations suggest that the turbulent cascade is driven by intermittent sheet-like structures in the ISM \citep{2022arXiv221010736D}.

If thin plasma sheets explain scintillation observations and are consistent with current understandings of the physics of turbulence in the ISM, might they not also explain ESEs? Here we propose a model for ESEs that arises naturally from the thin sheet picture which qualitatively explains several features of current ESE observations, including the complex frequency structure. The model we propose is a simple application of catastrophe theory at the density level of the lens description. That is, lensing by thin sheets can be effectively described by the projected density of the sheet onto a plane perpendicular to the line of sight. Mathematically, singularities in the projection map can be classified and described by a small set of elementary catastrophes \citep{thom1975structural}. Fold ($A_2$) catastrophes in corrugated plasma sheets have been proposed as an explanation for pulsar scintillation observations \citep{2006ApJ...640L.159G, 2014MNRAS.442.3338P, simard_predicting_2018}. Here we propose the next higher-order catastrophe, the $A_3$ cusp, as an explanation for ESEs. We call this framework ``doubly catastrophic" lensing, since catastrophe theory has long been applied to the theory of lensing to describe the magnification of sources near singularities in the lens map \citep{Nye}. In addition to describing the magnification as a network of catastrophes, here we describe the lens itself as a catastrophe. One of the advantages of such a framework, is that the elementary catastrophes are universal and described by a small number of unfolding parameters. Therefore, if correct, the effective lenses describing ESEs may be exceptionally simple in form, even if the physical plasma sheets are formed by complex physical processes. 

This paper is structured as follows. In Section~\ref{sec:model} we introduce our model for ESEs and discuss some of its qualitative features. In Section~\ref{sec:0834} we analyse in detail observations of an extreme scattering event in the pulsar PSR B0834+06 with our model. In Section~\ref{sec:scint} we discuss the possibility of using the doubly catastrophic lensing framework to explain both scintillation and ESEs, and in Section~\ref{sec:applications} we discuss potential applications of this framework.

\section{The $A_3$ lens}
\label{sec:model}

\begin{figure}
    \centering
    \includegraphics[width=\columnwidth]{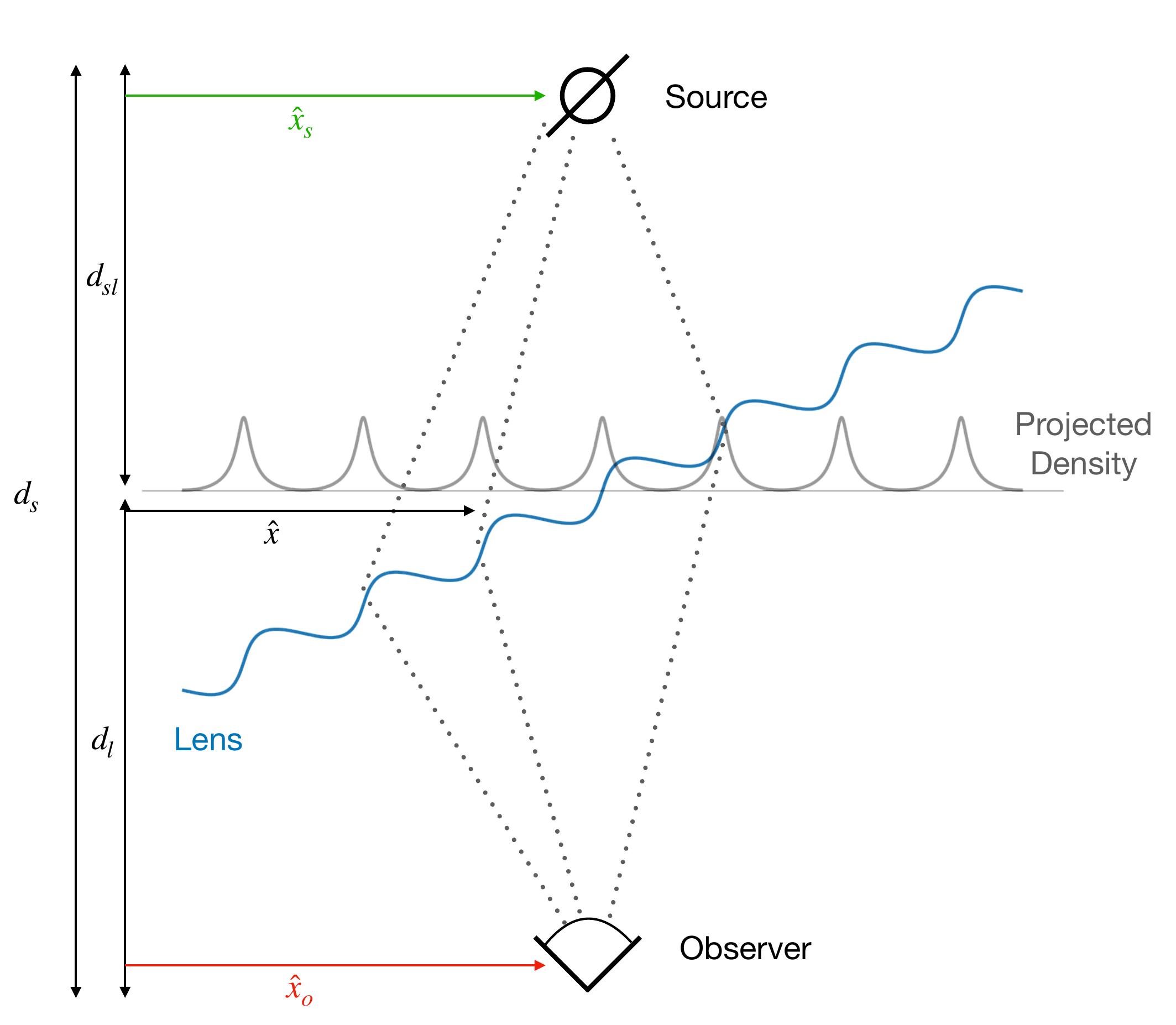}
    \caption{Diagram of a corrugated sheet lens. When the distances between the observer, source, and lens are large compared to the extent of the sheet along the line of sight (as in most astrophysical scenarios), the lensing is effectively described by the projected density (shown in gray) of the sheet onto the lens plane perpendicular to the line of sight. As the sheet is rotated to be more aligned with the line of sight, the peaks of the projected density become larger. The refraction of light due to the lens causes multi-path propagation from source to observer, shown by the grey, dashed lines.  }
    \label{fig:corr_diag}
\end{figure}

In geometric optics, the effect of a plasma lens localized to a single plane along the line of sight is determined by the lens equation
\begin{equation}
    \hat{\bm y} = \hat{\bm x} - \frac{\overline{d} c e^2}{2 m_e \epsilon_0 \omega^2} \nabla_{\hat{\bm x}} \Sigma_e (\hat{\bm x}),
    \label{eq:lenseq}
\end{equation}
where $\omega$ is the angular frequency of the light, $\hat{\bm y} = (\hat{\bm x}_s d_l + \hat{\bm x}_o d_{sl}) / d_s$ is a weighted average of the transverse displacement between the source and observer, $\overline{d} = d_{sl} d_l / d_{s}$ is an effective distance, and $\Sigma_e(\hat{\bm x})$ is the excess surface electron in the lens plane. The coordinates and distances involved are shown in Fig.~\ref{fig:corr_diag}. The lens equation determines a mapping between the source plane and the lens plane, determining the set of rays that connect the source and observer.

In astrophysical lensing, due to the vast distances between the source and the observer, it is often sufficient to treat lensing in this ``thin lens" approximation, where the lens is taken to be localized to a single plane perpendicular to the line of sight (the lens plane). The physical plasma that produces the lens effect is, of course, not a fully two-dimensional screen, but has some extent along the line of sight. As such, the surface density, $\Sigma_e$, is a projection of the actual density onto the lens plane:
\begin{equation}
    \Sigma_e (\hat{\bm x}) = \int \delta n_e (\hat{\bm x}, z) dz,
    \label{eq:projection}
\end{equation}
where $\delta n_e$ is the excess electron density. 

Fig.~\ref{fig:corr_diag} shows an example of this projection process. Consider a thin sheet with a periodic corrugation that is inclined by some angle with respect to the line of sight (the blue curve in Fig.~\ref{fig:corr_diag}). The surface density along the lens plane (the grey curve) is obtained by projecting the sheet onto a plane perpendicular to the line of sight. In particular, for an infinitely thin sheet with a shape given by $x = f(z)$, the projected density is proportional to $|\frac{dx}{dz}|^{-1}$. Thus, the density is formally infinite at singularities of the projection. 

Catastrophe theory describes the mathematics of such singularities. Powerfully, catastrophe theory shows that the topological structure of singularities must conform to a few fundamental forms (the ``elementary catastrophes") regardless of the precise details of the map in which the singularities arise. Catastrophe theory has already been used to great effect in lensing theory  to predict the magnification of a source near a lens' caustics without needing to know the precise details of the lens potential. Our proposal here is to extend this use of catastrophe theory to the density level. That is, we wish to describe not only the magnification via elementary catastrophes, but the lens potential itself. We expect these elementary forms are likely to appear in the projected plasma density, as catastrophes generically arise when projecting thin, sheet-like structures in the ISM along the line of sight. We will argue that by modelling the plasma structures responsible for ESEs by these catastrophes, it is possible to explain aspects of observations that have thus far been challenging to model. We call this framework of describing the projected plasma density by a network of caustics ``doubly catastrophic lensing", as catastrophes arise both in the magnification produced by the lens and in the lens potential, itself. 

\subsection{Modelling the $A_3$ lens}

\begin{figure}
    \centering
    \includegraphics[width=\columnwidth]{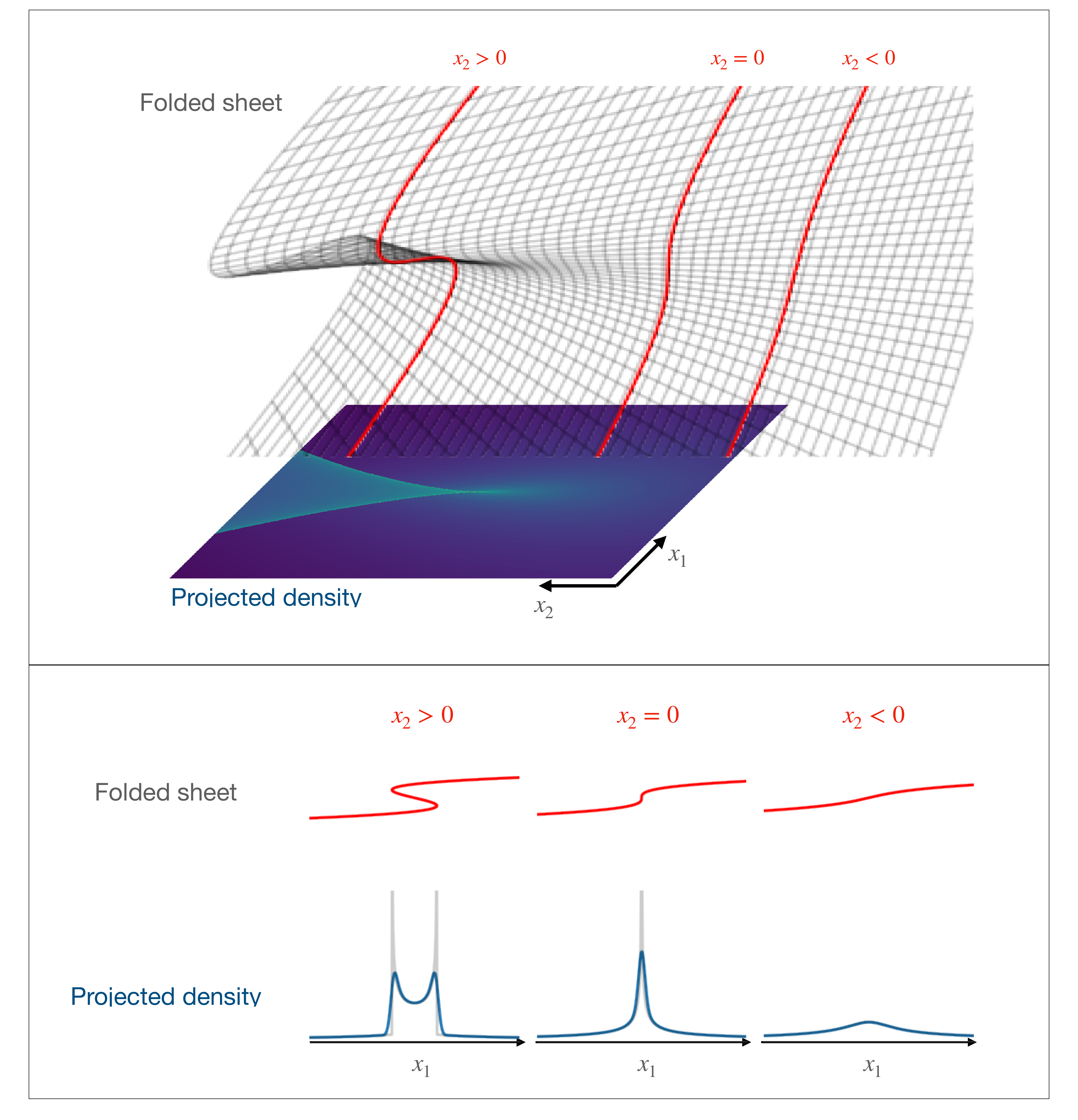}
    \caption{Diagram showing how the $A_3$ cusp catastrophe arises when a folded sheet is projected onto the lens plane. The grey mesh shows the physical sheet and the colour map below shows the density of the sheet when projected onto the lens plane. The cusp profile arises generically when two folds in the sheet come to an end. The variables $x_1$ and $x_2$ are the two unfolding parameters of the $A_3$ catastrophe and can be thought of as the physical coordinates in the lens plane up to some arbitrary scaling. The red lines show cross sections through the sheet for fixed $x_2$. In the bottom panel, we show the cross section through the physical sheet in red, and the projected density for an infinitely thin sheet below that in grey. The blue curves show the projected density for a sheet of finite thickness; the effect of which is to smooth out the sharply peaked grey curves. For $x_2 < 0$, the sheet is made up of two folds that converge at $x_2 = 0$ at the cusp point. For $x_2 > 0$, the two folds disappear as the sheet flattens out.}
    \label{fig:cusp_diag}
\end{figure}

Catastrophe theory is the mathematical classification of the stable singularities of continuous mappings. The power of catastrophe theory comes from the fact that the stable sinuglarities can be classified by a small number of canonical forms. In this paper we will focus on the $A_3$ catastrophe, a.k.a. the cusp catastrophe. The fold ($A_2$) and cusp catastrophes are the simplest of the elementary catastrophes. Lensing by a fold has been discussed elsewhere, and has been proposed as an explanation for pulsar scintillation \citep{2006ApJ...640L.159G, 2014MNRAS.442.3338P, simard_predicting_2018}. Here we propose lensing by an $A_3$ catastrophe as an explanation for ESEs in pulsars and quasars. The basic idea is shown in Fig.~\ref{fig:cusp_diag}; when the folds of a thin, folded sheet come to an end, they meet in a cusp. The sheet, when viewed under projection, forms an $A_3$ cusp density profile. The cusp is described by two unfolding parameters: $x_1$ and $x_2$. The bottom panel of Fig.~\ref{fig:cusp_diag} shows the cusp density as a function of $x_1$ for fixed $x_2$.

Concretely, we will use the canonical density of the $A_3$ catastrophe, which we will call $\phi_{A_3}(x_1, x_2)$, as the lens potential. The $A_3$ catastrophe is described by the Lagrangian map, $\xi: t \mapsto (x_1, x_2)$, defined by \citep{Nye}:
\begin{equation}
    t^3 - x_2 t + x_1 = 0.
    \label{eq:A3map}
\end{equation}
The $A_3$ catastrophe, or caustic, is the image of the singularities of this mapping, i.e. where the Jacobian of the map, $\xi'(t)$, vanishes. We will take as our lens potential the density field described by this mapping: 
\begin{equation}
    \phi_{A_3}(x_1, x_2) = \sum_{t_i = \xi^{-1}(x_1,x_2)} \frac{1}{\left| \xi'(t_i) \right|} = \sum_{t_i} | 3 t_i^2 - x_2 |^{-1}
    \label{eq:muA3}
\end{equation}
where the $t_i$ are the real solutions to Eq.~\ref{eq:A3map}. If we imagine the $t$ variable as the height of an infinitely thin sheet above the $(x_1,x_2)$-plane, then $\phi_{A_3}$ is the projection of sheet onto the plane, shown in  Fig.~\ref{fig:cusp_diag}. The catastrophe occurs when the denominator in the sum of Eq.~\ref{eq:muA3} is zero, i.e. the singularities of the projection. Defining $x_1^*$ and $x_2^*$ as the locations of the catastrophe, we find the relation
\begin{equation}
    x_1^* = \pm \frac{2 \sqrt{3}}{9} (x^*_2)^{3/2}.
    \label{eq:A3_cusploc}
\end{equation}
Thus the catastrophe opens up from a cusp at the origin into two folds given by the plus and minus sign of Eq.~\ref{eq:A3_cusploc}.

So far, we have been interpreting the quantities $x_1$ and $x_2$ as the dimensionless unfolding parameters of the cusp catastrophe. However, we want to interpret these as physical distances in the lens plane. In order to do so, while keeping $\phi_{A_3}$ dimensionless, we will define the dimensionless distances $\bf{x} = \bf{\hat{x}} / \ell$ and $\bf{y} = \bf{\hat{y}} / \ell$, normalized by some convenient physical scale $\ell$. Now, using $\phi_{A_3}$ as our lens potential, the lens equation which determines the mapping between the lens plane, $\bf{x}$, and the source plane, $\bf{y}$, is given by:
\begin{equation}
    {\bm y} = {\bm x} + \alpha \nabla \phi_{A_3}({\bm x}),
    \label{eq:dimless_lenseq}
\end{equation}
where we introduce the parameter $\alpha$ so that we may independently vary the amplitude of the lens potential. We will also introduce an additional parameter $\sigma$ and define
\begin{equation}
    \tilde{\phi}_{A_3}(x_1, x_2; \sigma) \equiv \phi_{A_3}(x_1, x_2) \star W(x_1; \sigma),
    \label{eq:smoothed_muA3}
\end{equation}
where $W(x_1; \sigma)$ is taken to be a simple Gaussian smoothing function with standard deviation $\sigma$. This smoothing is performed to remove the infinite densities that arise in the cusp catastrophe and represents the fact that the physical sheet that gives rise to the cusp has a finite thickness. 

Together, Eqs.~\ref{eq:dimless_lenseq} and \ref{eq:smoothed_muA3} provide a full description for the $A_3$ lens in geometric optics. However, to further simplify our analysis we will treat the lens as quasi-one-dimensional. That is, we will assume that the direction of the rays is only modified in the $x_1$ direction. In other words, we treat $x_2$ as a fixed parameter of the lens, and we only need to solve the one-dimensional lens equation:
\begin{equation}
    y_1 = x + \alpha \partial_x \tilde{\phi}_{A_3}(x; x_2).
    \label{eq:quasi1D_lenseq}
\end{equation}
The second lens equation is simply $y_2 = x_2$. This simplification is possible because the gradient of the smoothed potential, $\nabla \tilde{\phi}_{A_3}$, diverges transverse to the fold catastrophe as $\sigma \to 0$, but vanishes parallel to the fold. Thus, if we choose $\sigma \ll 1$, then the rays are effectively bent in only one direction: perpendicular to the fold. Then, when $y_1$ is far from the fold, the bending angle is almost entirely along the $x_1$-direction. Note that this is generically the case for any localized lens: when the source position, $\bf{y}$, is far from the lens, the bending angle is dominated by the gradient in the direction parallel to $\bf{y}$. Thus, the condition for our quasi-one-dimensional approximation to hold is that the source position is far from the catastrophe in units of $\sigma$, or, in other words, $|y_1 \pm x_1^*| / \sigma \gg 1$. By choosing small values of $\sigma$, we are generically in this regime, except very close to the catastrophes. However, note that while we can justify this assumption of quasi-one-dimensionality in this way, we can also regard this assumption as a phenomenological choice which we will show can explain certain aspects of observational data.

This simplification is possible because the derivative of $\tilde{\phi}_{A_3}$ is typically much larger in the $x_1$ direction than it is in the $x_2$ direction. While not strictly necessary here, this simplification will come in handy when we consider a multi-plane lens in Section~\ref{sec:0834}.

In this limit, the magnification due to the $A_3$ lens is given by
\begin{equation}
    \mu(y_1; x_2) = \sum_{x} | 1 + \alpha \partial^2_x \phi_{A_3}(x; x_2) |^{-1},
\end{equation}
where the sum is taken over solutions to the lens equation, Eq.~\ref{eq:quasi1D_lenseq}.

\begin{figure}
    \centering
    \includegraphics[width=\columnwidth]{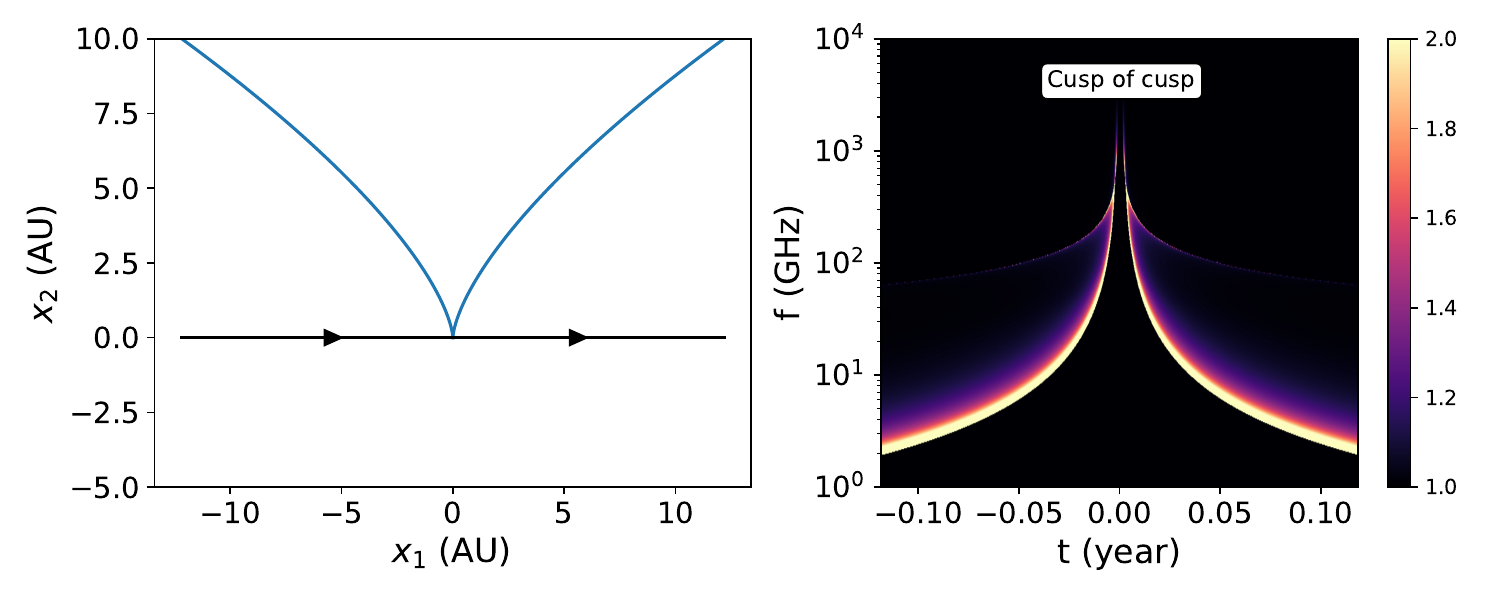}
    \caption{The right panel shows the dynamic spectrum (intensity as a function of time and frequency) one would observe for the source trajectory shown by the black arrows on the left. The blue curve on the left shows the caustics of the $A_3$ cusp profile, i.e., the points of maximum projected density shown in Fig.~\ref{fig:cusp_diag}. We have chosen the trajectory to just graze the cusp point. Here we are describing the lens potential as an cusp catastrophe, but cusps also generically arise in the dynamic spectrum. This is clearly seen in the right panel, where the bright peaks of the magnification converge towards a cusp at high frequencies: this is the titular "cusp of the cusp".}
    \label{fig:my_label}
\end{figure}

It turns out that for the $A_3$ potential, changes along the $x_2$ direction can be described by changes in the amplitude of the lens, and a re-scaling of the $x$ coordinate. This follows from the identity
\begin{equation}
    \alpha \phi_{A_3} (x, x_2 = \pm 1) = \phi_{A_3}\big(\alpha^{-3/2} x, x_2 = \pm \frac{1}{\alpha}\big),
\label{eq:alpha_rescale}
\end{equation}
or, equivalently,
\begin{equation}
    \phi_{A_3}(x, x_2) = \frac{1}{|x_2|} \phi_{A_3}\big(|x_2|^{-3/2} x, {\rm sign}(x_2)\big).
    \label{eq:x2_rescale}
\end{equation}
In other words, once a sign is chosen for $x_2$, one is free to re-scale the amplitude $\alpha$ and the $x$ coordinate so that $|x_2| = 1$. 

\begin{figure}
    \centering
    \includegraphics[width=\columnwidth]{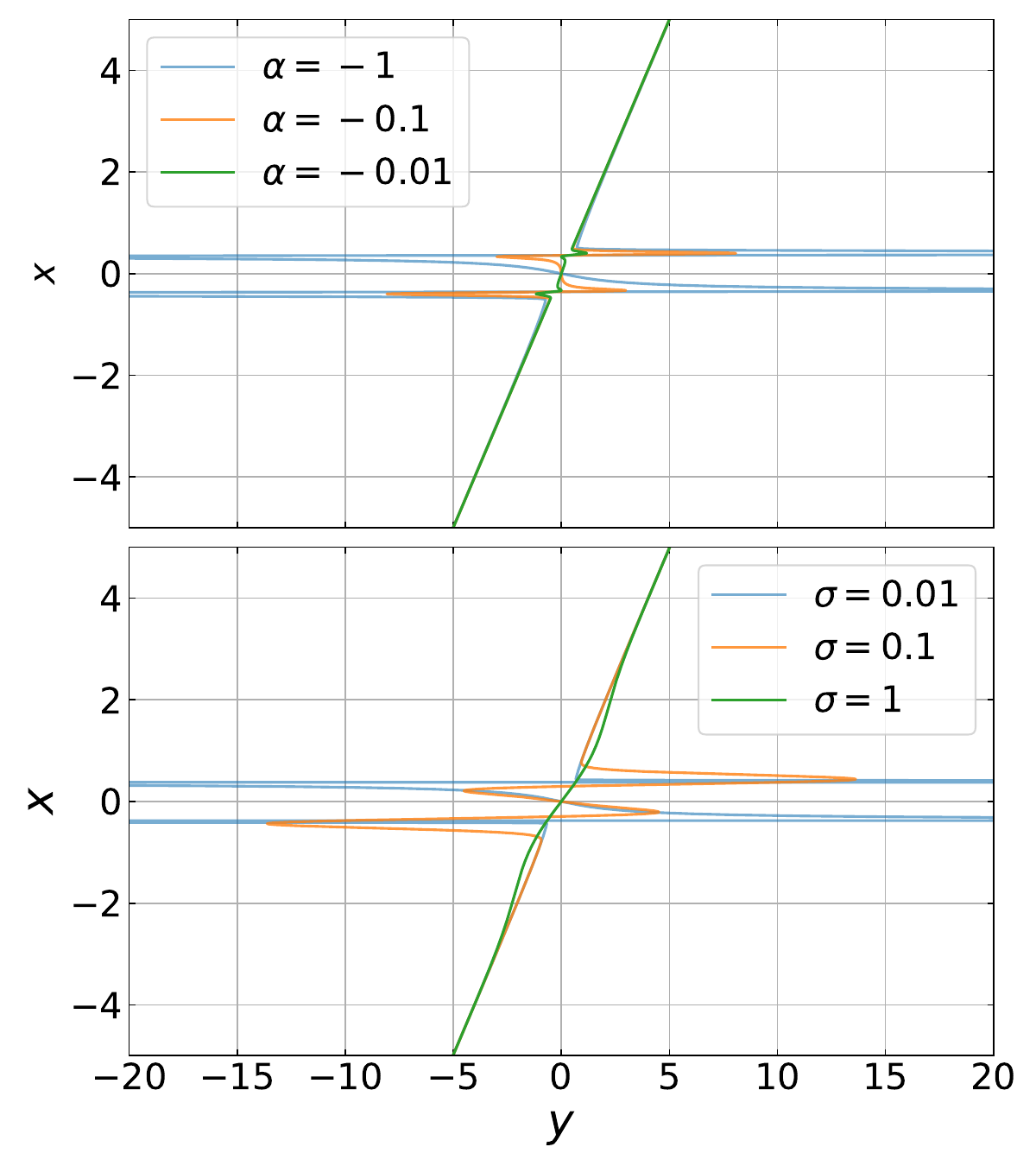}
    \caption{(Top) The lens map of the $A_3$ lens for fixed $x_2 = 1$, $\sigma = 0.03$ and varying $\alpha$. The effect of changing $\alpha$. (Bottom) The lens map for fixed $x_2 = 1$, $\alpha = -1$, and varying $\sigma$.}
    \label{fig:lensmap}
\end{figure}

The top panel of Fig.~\ref{fig:lensmap} shows the lens map described by Eq.~\ref{eq:quasi1D_lenseq} for fixed $x_2 = 1$ and $\sigma = 0.03$, for different values of $\alpha$. One may be interested in the location of the caustics that are formed by the lens (i.e. the turning points in the lens map). The location of these turning points depends on $\alpha$ and the smoothing scale $\sigma$, as the un-smoothed potential is formally infinite at certain points. Thus, the maximum amplitude of the smoothed potential depends strongly on $\sigma$. The bottom panel of Fig.~\ref{fig:lensmap} shows the lens map for fixed $x_2 = 1$ and $\alpha = -1$, for different values of $\sigma$. For $\alpha = -1$ and small $\sigma \ll 1$, there are six total caustics which form: four outer caustics which move out to infinity as $\sigma \to 0$, and two inner caustics which stay close to $y \sim 1$. The effect of increasing $\sigma$ is to increasingly smooth the caustics so that the outer caustics move in from infinity. Eventually, $\sigma$ becomes so large that some of the caustics disappear. For example, compare the green curve in the bottom panel of Fig.~\ref{fig:lensmap} to the orange and blue curves. Whereas the latter exhibit six caustics, the green curve only has two caustics close to the origin. Thus, not only are the position of the caustics dependent on $\sigma$, but also the number. In this work, we will restrict our attention to values of $\sigma \ll 1$, so that the number of caustics is the same as the number in the $\sigma \to 0$ limit. Otherwise, one is essentially smoothing away the unique features of the $A_3$ lens, erasing the predictive power of this model.

Now, it is straightforward to compute how the location of the outermost caustics scale with $\sigma$ and $\alpha$, for $\sigma \ll 1$. Roughly, the location of the outermost caustic is given by $y^*_1 \sim {\rm max}\{\alpha \partial_x \phi \}$. For the un-smoothed lens, the maximum derivative of the lens potential is infinite. For the smoothed lens, the maximum value of the derivative is nevertheless attained close to where the unsmoothed lens diverges. Since the lens potential near the divergence is described by a fold catastrophe, the lens potential is given by $\phi(x; x_2) \propto x^{-1/2}$ on one side of the divergence and zero on the other side. Thus, near the divergence, the smoothed derivative is given by $\partial_x \phi(x; x_2) \propto \phi(x; x_2) \star \partial_x W(x; \sigma)$. It follows from this that the location of the outermost caustic scales as
\begin{equation}
    y^*_1 \sim \alpha \sigma^{-3/2}.
    \label{eq:ystar}
\end{equation}
Knowing the location of the outermost caustic will be useful later when we are trying to infer the lens parameters from observed time delays in Section~\ref{sec:0834}.

\begin{figure}
    \centering
    \includegraphics[width=\columnwidth]{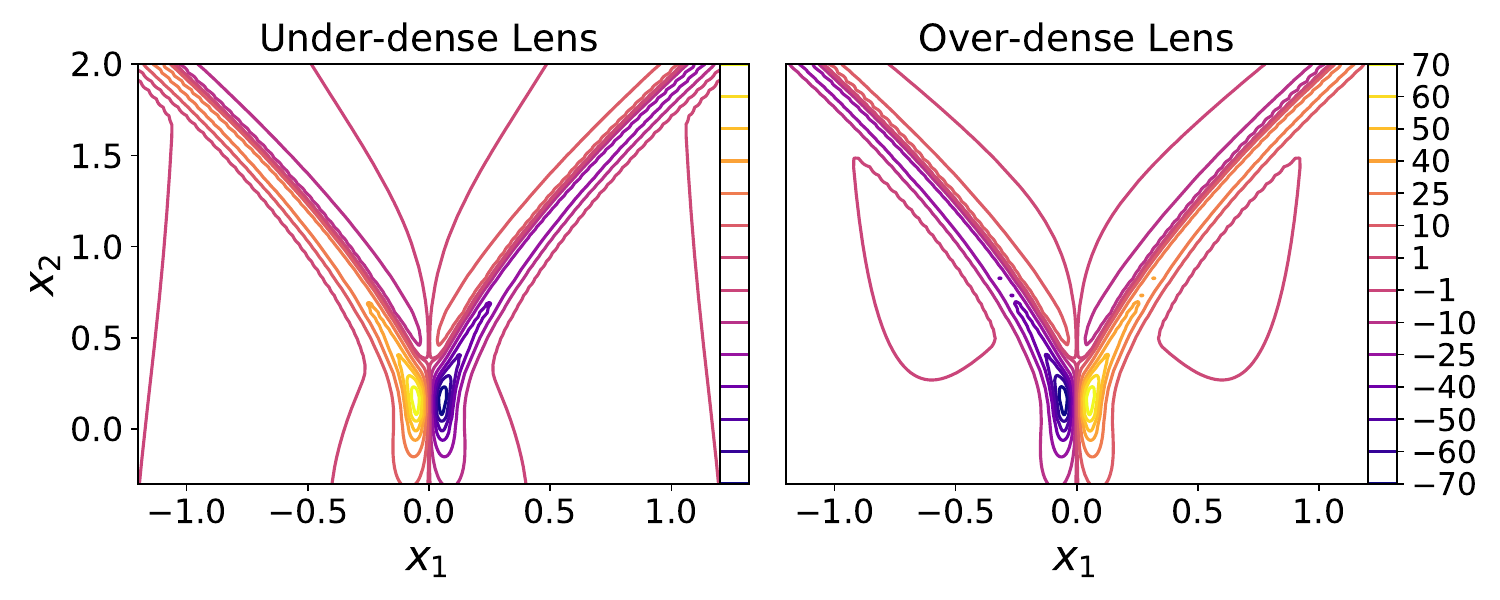}
    \caption{The level curves of the lens map, Eq.~\ref{eq:quasi1D_lenseq}, as a function of the unfolding parameters $x_1$ and $x_2$, for fixed $\alpha = \pm 1$ and $\sigma = 0.03$.}
    \label{fig:contours}
\end{figure}

Fig.~\ref{fig:contours} shows the level curves of the lens map (Eq.~\ref{eq:quasi1D_lenseq}), as a function of the unfolding parameters, $x_1$ and $x_2$, for fixed $\alpha = \pm1$ and $\sigma = 0.03$, where the sign of alpha determines whether the lens is under- or over-dense (note that because of the scaling relations shown in Eqs.~\ref{eq:alpha_rescale} and \ref{eq:x2_rescale} varying either $\alpha$ or $x_2$, while holding the other fixed, covers the entirety of the parameter space of the lens). Fig.~\ref{fig:contours} tells us where a given source position $y_1$ gets mapped to in the lens plane. That is, consider $x_2$ to be some fixed value. Then we can read off the image positions in $x_1$ for a given value of $y_1$ by looking at the $x_1$ value the contour associated with $y_1$ reaches for that value of $x_2$. A peculiar feature of the $A_3$ lens is that for a fixed source position, for a small region about $x_2 = 0$, the distance between the image positions in $x_1$ increases as $x_2$ decreases to zero. This is in contrast to large values of $x_2$, for which a decrease in $x_2$ leads to the images moving closer together. This feature will lead to the characteristic hockey-stick shape shown later in Figs.~\ref{fig:datasims} and \ref{fig:div_vs_conv}.

\begin{figure}
    \centering
    \includegraphics[width=\columnwidth]{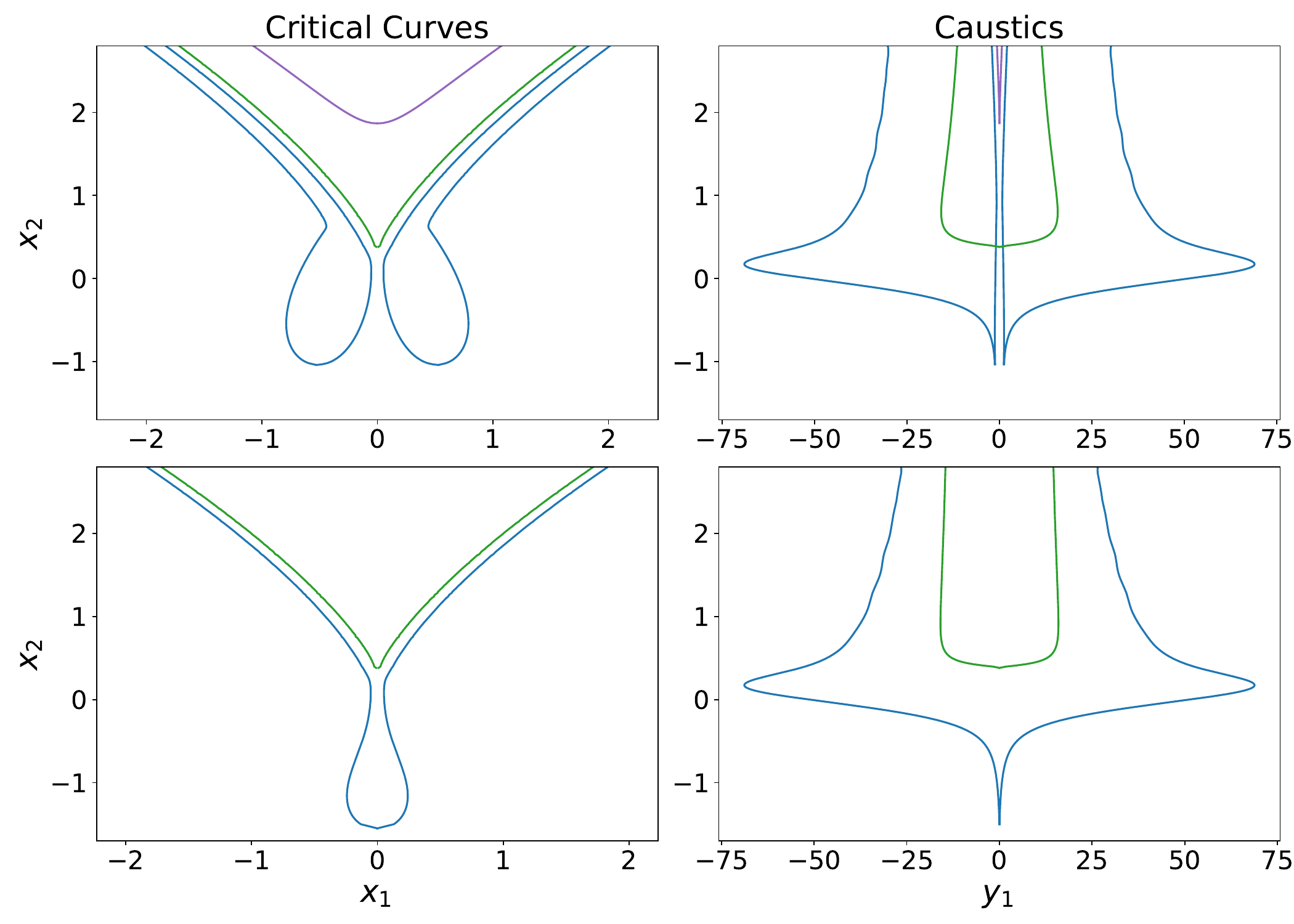}
    \caption{The critical curves and caustics of the $A_3$ lens for fixed $|\alpha| = 1$ and $\sigma = 0.03$. The top row shows the results for the over-dense lens, $\alpha < 0$, and the bottom row shows the results for the under-dense lens, $\alpha > 0$. }
    \label{fig:crit_caust}
\end{figure}

\begin{figure}
    \centering
    \includegraphics[width=\columnwidth]{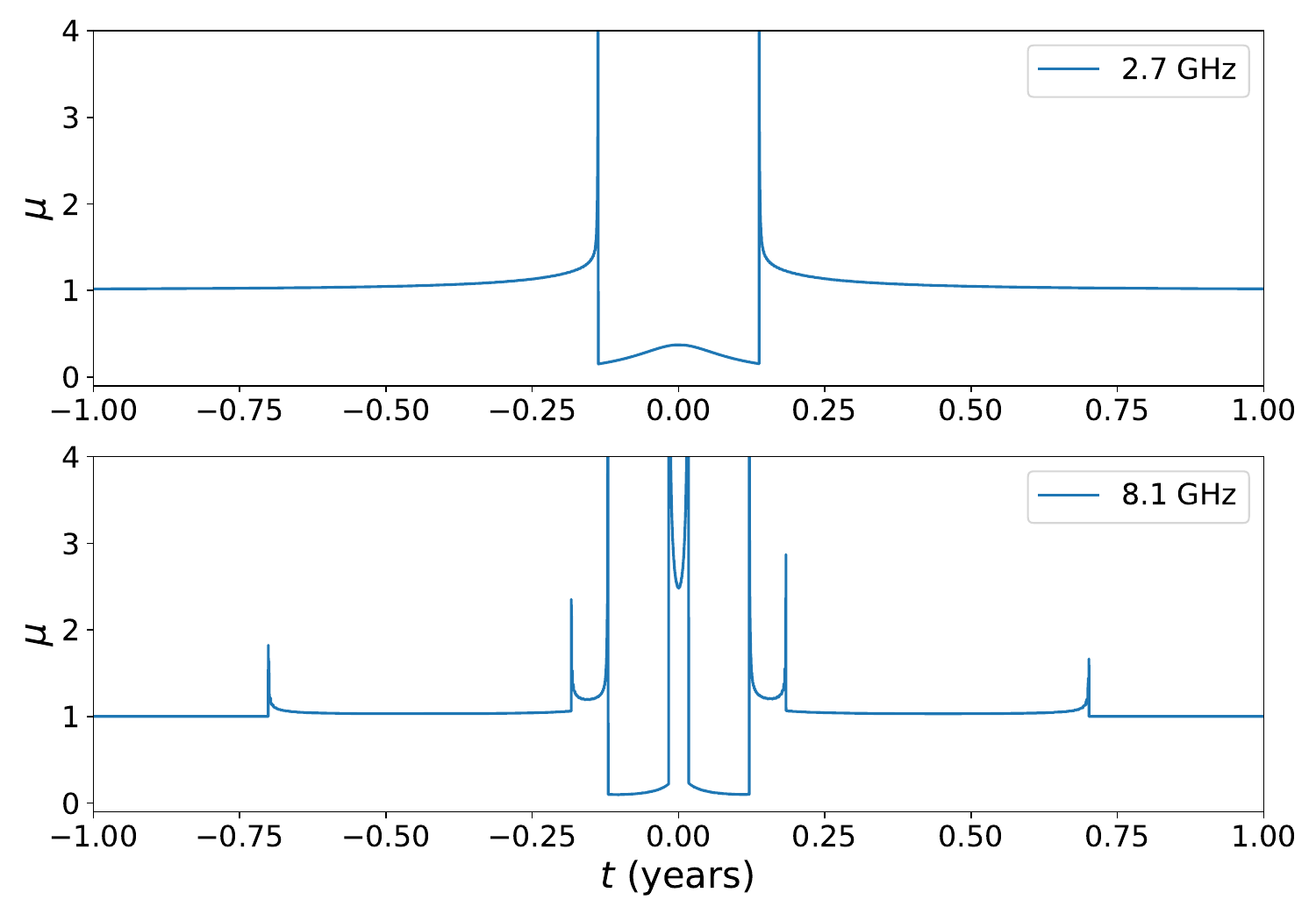}
    \caption{An example light curve for two frequency bands, $f = 2.7\,{\rm GHz}$ and $f = 8.1\,{\rm GHz}$, for the $A_3$ lens. The light curve is computed for $d_l = 1\,{\rm kpc}$, $d_s = 1\,{\rm Gpc}$, and fixed $x_2 = 10\,{\rm AU}$. We choose $\alpha = -0.3$ at $f = 2.7\,{\rm GHz}$, and a smoothing scale $\sigma_x = 0.03$, which for the chosen parameters corresponds to a peak electron surface density of $\Sigma_e \approx 0.02\,{\rm pc}\,{\rm cm}^{-2}$. The source position as a function of time is given by $y = v t$ where $v = 200\,{\rm km}\,{\rm s}^{-1}$.}
    \label{fig:lc}
\end{figure}

Fig.~\ref{fig:crit_caust} shows the critical curve and caustic structures that arise due to magnification by the $A_3$ lens. The top row shows the results for the over-dense lens ($\alpha < 0$), and the bottom row shows the under-dense lens ($\alpha > 0$). The light curve that arises as one changes impact parameter, $y_1$, (i.e. as the source moves relative to the lens), are effectively entirely determined by the number and location of the caustics. When $\sigma \ll 1$ and $\alpha \gtrsim 1$, the caustics tend to be located far from the axis; for example, the outermost caustic is located at $y^*_1 \gg 1$. This means that in between the caustics the slope of the inverse lens map is large, meaning that the total magnification is close to one (see, for example, the blue curve in Fig.~\ref{fig:lensmap}, where the inverse lens map, $x(y)$, is effectively flat for most of the region between the caustics). The result is that the light curve as the source moves relative to the lens is close to unity except at the caustics where the magnification suddenly diverges. Fig.~\ref{fig:lc} shows an example light curve for $\alpha = -0.3$, $\sigma = 0.03$. The lens is taken to be of size $\ell = 10\,{\rm AU}$ and moving with velocity $v = 200\,{\rm km}{\rm s}^{-1}$ relative to the lens. Here $\ell$ is simply a normalization parameter to convert the dimensionless distances to dimensionful ones. Its relation to the lens size is clarified in Section~\ref{sec:0834}. The parameters are chosen to match the $f = 2.7\,{\rm GHz}$ light curve of the original ESE observation presented in \citet{Fiedler1987}. Since for plasma lensing $\alpha \propto f^{-2}$, we can compute the light curve for multiple frequency bands. Computing the light curve for $f = 8.1\,{\rm GHz}$ (shown in the bottom panel of Fig.~\ref{fig:lc}), we see that, at high frequencies, multiple caustics in the light curve become apparent. It is not the case that these magnification caustics appear because new features in the lens become resolved as the frequency increases. Rather, as is clear from Eq.~\ref{eq:ystar} and shown in Fig.~\ref{fig:lensmap}, as $\alpha$ decreases, the location of the caustics move in from infinity. In other words, if one were to compute the light curve over many frequencies between $f = 2.7$ and $8.1\,{\rm GHz}$, one would see the intensity spikes move continuously inward from infinity, as opposed to appearing spontaneously. This latter situation would occur if the additional caustics at high-frequency were due to unknown substructure in the lens which becomes resolved at high frequency. Note, however, this prediction of caustics moving continuously in from infinity is not unique to our model, but is generic to one-dimensional, smooth refractive lenses. Nevertheless, the high-frequency light curve shown in Fig.~\ref{fig:lc} shares qualitative similarities to the high-frequency observation of the original ESE in \citet{Fiedler1987}, which has otherwise been challenging to accommodate with other refractive lens models. While we have not undertaken a quantitative best-fit analysis of the observations with our model, we argue that the $A_3$ lens naturally explains the appearance of multiple magnification caustics at high frequencies without the need to invoke unknown substructure as is often done in many attempts to model ESEs.  

\subsection{A physical picture}


We will now briefly discuss a possible physical origin for $A_3$ lenses. First, however, we will note that such lenses will generically arise for any lens that can be described by a thin-lens approximation. This is because the mathematics of catastrophe theory requires singularities of projection maps to take on a small number of generic forms (the elementary catastrophes). Given the vast distances involved in astrophysical lensing, all but the most extended lenses will be adequately described by the thin-lens approximation. Thus, the presence of $A_3$ lenses does not strongly depend on a particular physical model; $A_3$ lenses should arise generically in most models of the ISM. The question is not whether or not there are $A_3$ lenses, but whether or not the $A_3$ lenses predicted by a given physical model can explain the observed properties of ESEs. 

In order for $A_3$ lenses to be a reasonable candidate to explain observed ESEs, we require that the transverse physical size of the lenses (projected onto the plane of the sky) be roughly on the order of $1\,{\rm AU}$, as this is the typical physical scale that can be inferred from ESE observations. We also require that the thickness of the sheet that is projected to produce these lenses be much less than $1\,{\rm AU}$. In other words, we require $\sigma \ll 1$. While, in principle, there is nothing stopping us from modelling lensing events with a highly smoothed $A_3$ lens, when the smoothing scale is large ($\sigma \gtrsim 1$), the unique features of the cusp become washed out, lessening the explanatory power of such a model. A third requirement is that whatever physical process causes the $A_3$ lenses, the lenses must persist on a timescale of months to years in order to match the timescale of ESE observations. 

A physical picture for corrugated sheets that satisfies these properties has already been proposed \citep{2006ApJ...640L.159G} and has been suggested as a potential explanation for pulsar scintillation \citep{2014MNRAS.442.3338P, simard_predicting_2018}. The basic idea, which we will summarize here, is that magnetic re-connection sheets in the ISM (boundaries between oppositely oriented magnetic field lines) sustain plasma current sheets. Ducted waves, driven by the tension produced by the magnetic fields, propagate through the current sheets, forming a corrugated pattern in the plasma sheet. That is, the sheets are roughly uniformly dense, but instead of being perfectly flat sheets, form ridges and valleys which are highly elongated in one direction, much like corrugated cardboard. When these current sheets are close to aligned with the line of sight, these corrugated patterns produce the fold ($A_2$) and cusp ($A_3$) lenses when projected onto the lens plane (see e.g. Fig.~\ref{fig:corr_diag}). While the ducted waves propagate at the speed of sound in the plasma ($c_s \sim 10 T^{1/2}_4 {\rm km}\,{\rm s}^{-1}$, $T_4 \equiv T / 10^4\,{\rm K}$), when the sheet is aligned with the line of sight, the transverse speed of the waves projected onto the lens plane can be made arbitrarily small, depending on the degree of alignment. This leads to the long timescales over which the ESE lens structures are observed to persist \cite{Liu2016}. These magnetic re-connection sheets are also predicted to arise on the spatial scales required to explain ESE observations. While the energetic processes that stir the ISM (e.g. supernovae, ionization fronts, spiral density waves, etc.) are typically short lived and occur on parsec scales or larger, magneto-hydrodynamic simulations of turbulent dynamos demonstrate that stable magnetic re-connection sheets may occur well below the stirring scale, and, in particular, on the several AU scale required by ESEs. Moreover, these sheets are indeed predicted to be ``thin" relative to the transverse AU scale \citep{2006ApJ...640L.159G,2014MNRAS.442.3338P}.

It remains to be seen how realistic such a model of the small-scale ISM is. Realistic, high-resolution simulations are needed. Recent magnetohydrodynamic simulations of the turbulent ISM have revealed the ubiquity of thin, filamentary-like structures on small scales intermittently permeating the diffuse medium \citep{2022arXiv221010736D, 2022arXiv221106434F}. However, the resolution of these simulations are typically at much larger scales than the $\sim$AU scales we require to explain ESEs. Nevertheless, these recent simulations give some confidence that these thin, intermittent structures may plausibly exist.  However, we stress again that one of the strengths of the doubly catastrophic lensing framework is that it does not depend crucially on the details of the underlying physical model. We have summarized this particular model to give an outline of a plausible, but not necessary scenario that could give rise to the kinds of lenses we are considering. 

\section{A double lensing event}
\label{sec:0834}
\begin{figure}
    \centering
    \includegraphics[width=\columnwidth]{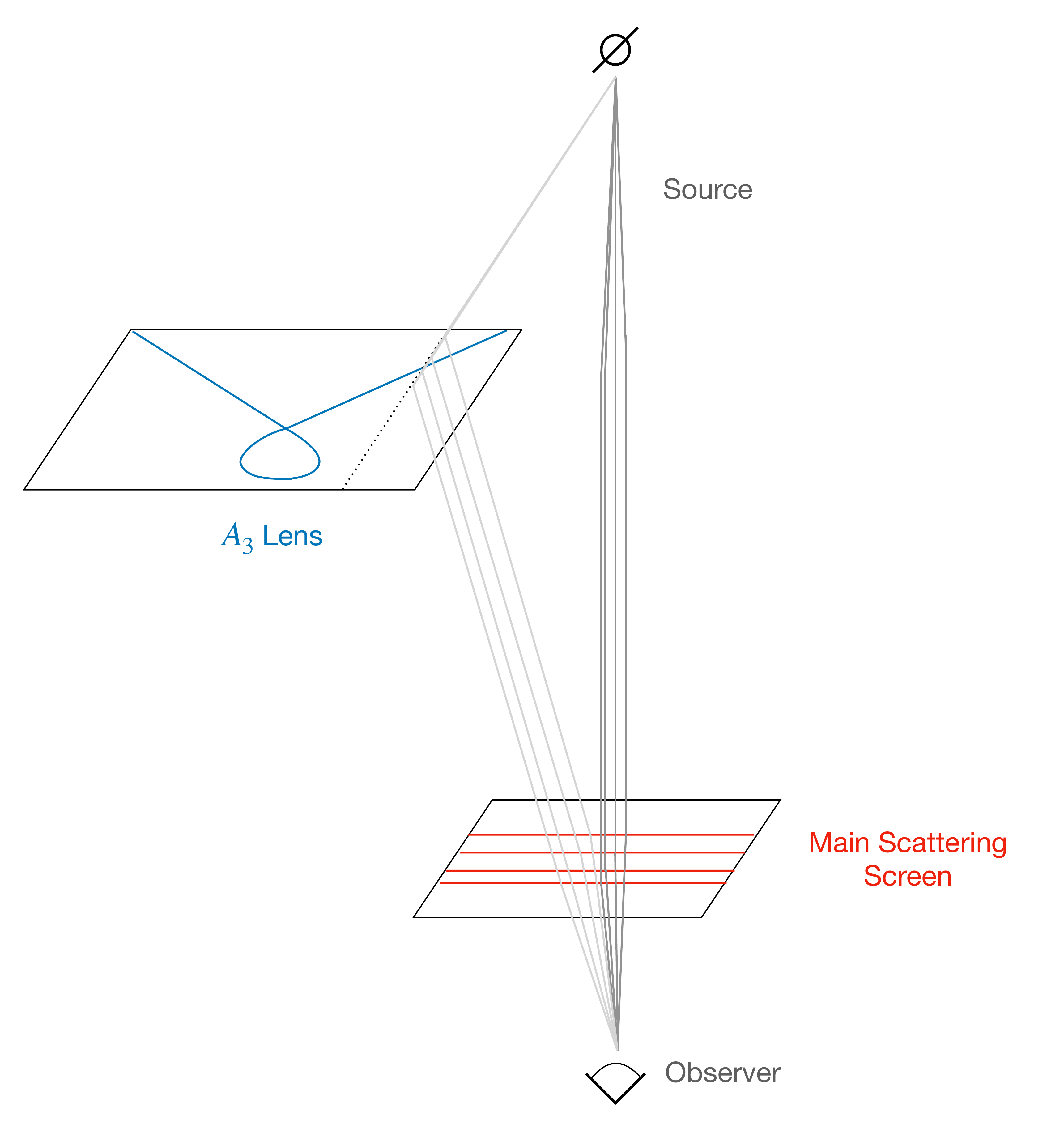}
    \caption{Diagram showing the geometry of the extreme scattering event observed in PSR B0834+06. The pulsar is first scattered by the ESE lens, which we propose is an $A_3$ lens, and is then scattered by the primary scattering screen which results in the parabolic arc that is ubiquitous in scintillation observations.}
    \label{fig:double_diag}
\end{figure}

\begin{figure}
    \centering
    \includegraphics[width=\columnwidth]{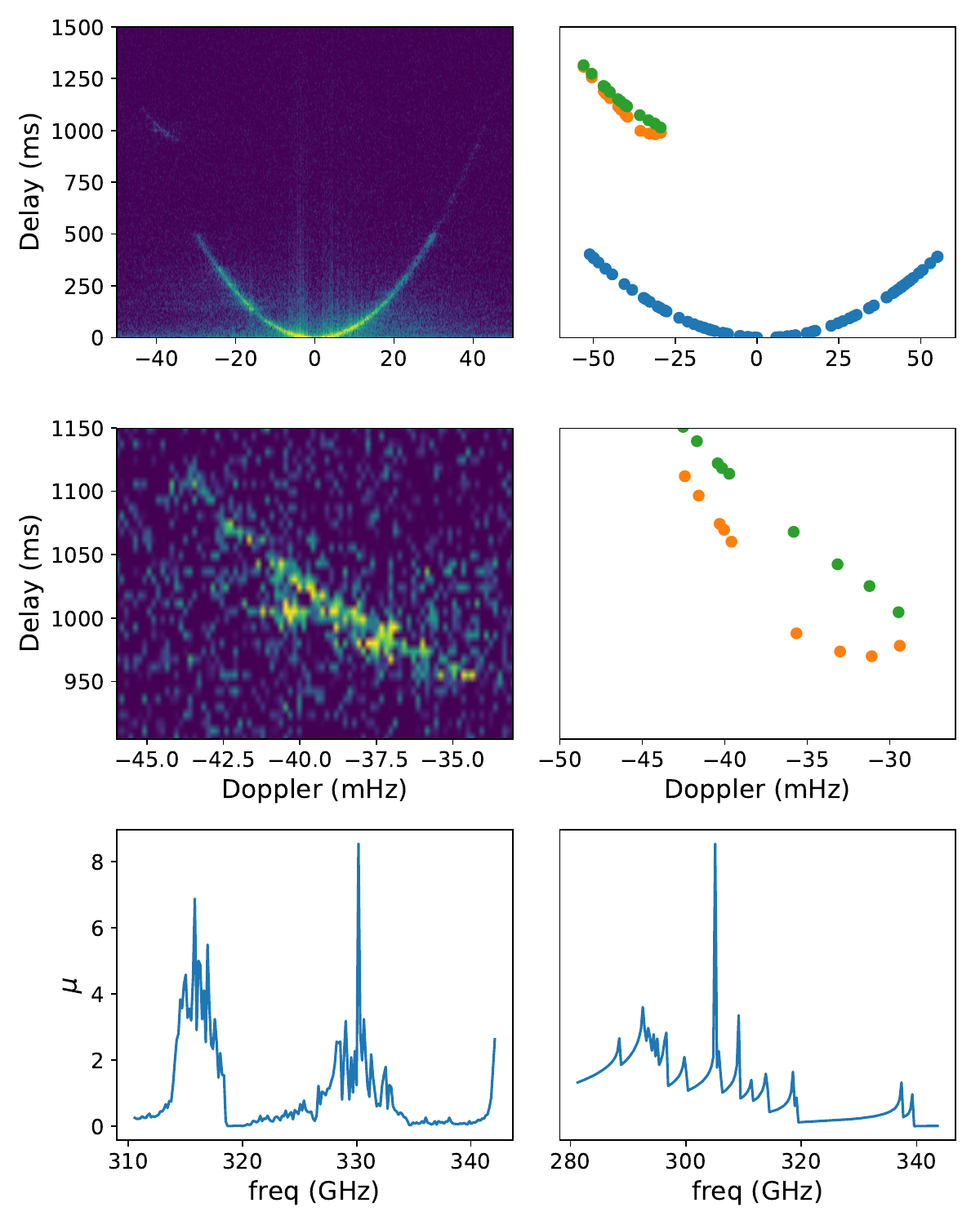}
    \caption{The left column shows data for an observation of PSR B0834+06. The top panel shows the conjugate wave-field of the observation in Doppler-delay space. The middle panel is the same as the top panel, zoomed in on the millisecond-delay feature associated with the ESE lens. The bottom panel shows the sum of the power in the conjugate wave-field of the millisecond wave-field as a function of frequency, normalized to unity when the signal falls below the noise threshold. This is taken as a proxy for the magnification induced by the lens. The top and middle panel of the right column shows the location of the images in Doppler-delay space for our double lensing model with an $A_3$ lens plus a primary scattering screen. We choose the dimensionless parameters $\alpha = 0.7$, $\sigma = 0.05$, $\nu = 31,000$, $D = 5$, $y_1 = 0$, $y_2 = 11$, and the direction of the velocity to be $\frac{\partial {\bm v}}{\partial t} \parallel [1, 0.1]$. The locations of the images on the scattering screen are chosen to be distributed uniformly over $z_1 \in [-16, 16]$. To convert to dimensionful parameters, we choose $f = 311\,{\rm MHz}$, $d_{01} = 389\,{\rm pc}$, $d_{02} = 415\,{\rm pc}$, and $d_{03} = 620\,{\rm pc}$ and a physical scale of the lens $\ell = 1\,{\rm AU}$. The magnitude of the velocity is chosen to be $23\,{\rm km}\,{\rm s}^{-1}$. These parameters correspond to an amplitude of the plasma under-density of $\Sigma^*_2 \sim 0.0001\,{\rm pc}\,{\rm cm}^{-3}$ (note that the maximum density at the fold is about an order of magnitude higher than this). Since we choose the two lenses to be perpendicular to each other, it is well-defined to identify each image with one of the three images produced by the $A_3$ lens: distinguished in the figure by the green, orange, and blue colours. The bottom right panel shows the total flux of the green and orange images (the images associated with the millisecond feature) as a function of frequency.}
    \label{fig:datasims}
\end{figure}

Now that we have outlined the details of the $A_3$ lens, we will turn to a
particular lensing event of interest in the pulsar PSR B0834+06.This
event has been a particularly fruitful object of study since its
observation in 2005 (Brisken et al. 2010) with the William E.Gordon
Telescope at the Arecibo Observatory, whose data we use here. The data
was taken in a 32 MHz band centred at 316.5 MHz over the course of $\sim
2$ hours. The dynamic spectrum was created using 5s integrations with
$\sim 0.25~\rm{kHz}$ channels. In order to collapse the inverted arclets
into single points to more easily identify images, we use the conjugate
wavefield produced by \cite{2022MNRAS.510.4573B} using phase retrieval
techniques to recover the electric field from the dynamic spectrum. The
conjugate wave-field (the top-left panel of Fig.~\ref{fig:datasims}) shows the main parabolic arc that is ubiquitous in
scintillation observations, in addition to a peculiar island of power
located at a delay of roughly $1\,{\rm ms}$ and Doppler shift of
$-40\,{\rm mHz}$. We will refer to this feature as the ``millisecond
feature". \citet{2022arXiv220806884Z} use observations over four epochs
in roughly fifteen-day intervals to demonstrate that this event is best
explained by a double lens system. That is, they argue the pulsar is
lensed by a main scattering screen, producing the primary scintillation
arc, and a second lens producing the millisecond features (a schematic
of this is shown in Fig,~\ref{fig:double_diag}).
\citet{2022arXiv220806884Z} use novel phase retrieval techniques to
infer the distances to the two screens, as well as the angular position
of the many images produced by this lensing system. From this they argue
that the secondary lens associated with the millisecond feature has
similar properties to the plasma structures responsible for ESEs. In
particular, its persistence over the more-than-month-long observation
and large bending angles ($\theta \approx 83\,{\rm mas}$) are consistent
with other ESE observations.

Here we will consider the possibility that this millisecond feature is actually produced by an $A_3$ lens. In order to do this, we first need to introduce the double lensing formalism. For a multi-plane lens, the induced phase along a particular path from source to observer is given by \cite{}
\begin{equation}
    S = \omega \sum_{i=1}^{N} \frac{d_{0i}d_{0 i+1} }{c d_{i i+1}} \Big[\frac{1}{2} ({\bm \theta_{i+1}} -  {\bm \theta_i})^2 + \frac{d_{i i+1} d_{0 n+1}}{d_{0 i+1} d_{i n+1}} \hat{\phi}_i ({\bm \theta_i})\Big],
\end{equation}
where $d_{ij}$ is the distance from the $i^{\rm th}$ lens plane to the $j^{\rm th}$ lens plane, and $i = 0, N$ refer to the observer and source, respectively. The angular coordinates associated with the $i^{\rm th}$ lens plane is given by ${\bm \theta}_i$ and $\hat{\phi}_i$ is the $i^{\rm th}$ lens potential. 

For a two-plane system, we can re-write this phase in terms of dimensionless parameters as follows:
\begin{equation}
    S = \nu \Big[\frac{\overline{d}_2}{\overline{d}_1} \big[ \frac{1}{2}({\bm x} - {\bm z})^2 + \rho \phi_1({\bm z}) \big] + \frac{1}{2} ({\bm x} - {\bm y})^2 + \alpha \phi_2 ({\bm x})\Big],
    \label{eq:S}
\end{equation}
where we have defined the co-ordinates ${\bm z} \equiv {\bm \theta_2} d_{01}/\ell$, ${\bm x} \equiv {\bm \theta_2} d_{02}/\ell$, and ${\bm y} \equiv {\bm \theta_3} d_{02} / \ell$ to be the physical distance in the respective lens/source plane, re-scaled by some physical scale $\ell$. For our purposes, we will take the first lens plane to be the main scattering screen, and the second lens plane to be the $A_3$ lens. It is, therefore, convenient to choose $\ell$ to be a physical scale associated with the $A_3$ lens. The barred distances are combined distances given by $\overline{d}_1 = d_{12} d_{02} / d_{01}$ and $\overline{d}_2 = d_{23} d_{02} / d_{03}$. The phase is multiplied by an overall factor of $\nu = \omega \ell^2 / \overline{d}_2 c$ and the amplitudes of the lens potentials are given by
\begin{align}
    \alpha &= \frac{\overline{d}_2 e^2 \Sigma^*_2}{2 m_e \epsilon_0 \omega^2 \ell^2}, \\
    \rho &= \frac{d_{12} d_{03}}{d_{02} d_{13}} \frac{\overline{d}_1 e^2 \Sigma^*_1}{2 m_e \epsilon_0 \omega^2 \ell^2}
    \label{eq:rho},
\end{align}
where $\Sigma^*_1$ and $\Sigma^*_2$ set the amplitude of the projected electron density of the lenses. For the $A_3$ lens, the maximum electron column density can be made arbitrarily large as $\sigma \to 0$. Thus, the actual electron column density will depend on the details of the lens. The parameter $\Sigma^*$ simply sets the approximate magnitude of the projected density.

Now, note also that we have defined $\nu = \omega \ell^2 / \overline{d}_2 c$, but we could just have easily defined $\nu = \omega \ell^2 / \overline{d}_1 c$, factoring out an overall factor of $\overline{d}_1$ as opposed to $\overline{d}_2$; however, for our purposes it is convenient to treat the $A_3$ lens as the primary lens, absorbing the geometric factors that appear in Eq.~\ref{eq:rho} into $\rho$ rather than $\alpha$. This, however, is purely a choice of convention, as the image locations and magnifications in geometric optics do not depend on the overall factor $\nu$.

The locations of the geometric images are given by the lens equations, $\nabla_{{\bm x}/ {\bm z}} S = 0$, which are:
\begin{align}
   & D (x_1 - z_1) + (x_1 - y_1) + \frac{d \phi_2}{d x_1} = 0, \\
   & D (x_2 - z_2) + (x_2 - y_2) + \frac{d \phi_2}{d x_2} = 0, \\
   & x_1 - z_1 + \frac{d \phi_1}{d z_1} = 0, \\
   & x_2 - z_2 + \frac{d \phi_1}{d z_2} = 0,
\end{align}
where we have defined the ratio $D \equiv \overline{d}_2 / \overline{d}_1$. 

Now, for the sake of simplicity, we will assume that the lenses are both highly anisotropic (i.e. one-dimensional) and that they are perpendicular to each other. That is, we will assume $\phi_2(x_1, x_2) = \phi_2(x_2)$ and $\phi_1(z_1, z_2) = \phi_1(z_1)$. In this way, the lens equations simplify to
\begin{align}
   & x_2 - y_2 + \frac{d \phi_2}{d x_2} = 0,
   \label{eq:x2}\\
   & x_1 - z_1 + \frac{d \phi_1}{d z_1} = 0,
   \label{eq:z1}\\
   & x_1 - \frac{y_1 + D z_1}{D + 1}  = 0, \label{eq:x1}\\
   & x_2 - z_2  = 0.
   \label{eq:z2}
\end{align}
It is convenient to do this because the result is that the two lenses act independently from each other; that is, we can solve the lens equations for $x_2$ and $z_1$ co-ordinates of the images independently using Eqs.~\ref{eq:x2} and \ref{eq:z1}, respectively, which then directly give us the $x_1$ and $z_2$ co-ordinates through Eqs.~\ref{eq:x1} and \ref{eq:z2}. In general, this simple separation of the lens equations into independent equations is not possible since the two lenses will not generically be perfectly perpendicular to each other. However, for our purposes in this work, we are primarily interested in the qualitative aspects of the $A_3$ lens, as opposed to a precise quantitative comparison, and \citet{2022arXiv220806884Z} show that the two lenses are, indeed, roughly perpendicular to each other.

In order to simulate the lensing event of PSR\,B0834+06, we will take $\phi_2(x_2) = \phi_{A_3} (x_2; x_1)$: the $A_3$ lens. Again, we stress that we are treating the $A_3$ lens as a quasi-one-dimensional lens, where the second co-ordinate $x_1$ is treated as a lens parameter. For the main scattering screen, instead of specifying a lens potential, we will simply specify a set of co-ordinates, $z_1$, fixing the location (in the $z_1$ direction) on the main scattering screen the rays must pass through. The goal of this is to re-produce the main scintillation arc seen in the conjugate wave-field without having to over-commit ourselves, as it were, to a particular scintillation model for the main screen. 

Once we have the location of the images by solving the lens equations, it is straightforward to compute where the images should appear in Doppler-delay space. The group delay of the images is given by
\begin{align}
\begin{split}
    \tau = \frac{\partial S}{\partial \omega} &= \frac{\nu}{\omega} \Big[D \big[ \frac{1}{2}({\bm x} - {\bm z})^2 - \rho \phi_1({\bm z}) \big] + \frac{1}{2} ({\bm x} - {\bm y})^2 - \alpha \phi_2 ({\bm x})\Big], \\
    &\approx \frac{\nu}{\omega} \Big[\frac{D}{2}({\bm x} - {\bm z})^2  + \frac{1}{2} ({\bm x} - {\bm y})^2 - \alpha \phi_2 ({\bm x})\Big].
    \label{eq:delay}
\end{split}
\end{align}
Note that the dispersive terms appear with a relative minus sign compared to Eq.~\ref{eq:S} since for plasma lensing the amplitudes of the lens potential have a frequency dependence $\alpha, \rho \sim \omega^{-2}$. We drop the dispersive term related to the main scattering screen as we have not specified the lens potential $\phi_1$. We assume that the delay from the main scattering screen is primarily geometric. 

The Doppler shift of the images is given by $f_D = \frac{d {\bm y}}{dt} \cdot \nabla_{\bm y} S$, which can be computed from the following:
\begin{align}
\begin{split}
    \frac{\partial S}{\partial y_1} &\approx -\frac{\nu D}{D + 1} (z_1 - y_1), \\
    \frac{\partial S}{\partial y_2} &\approx -\nu (x_2 - y_2).
    \label{eq:doppler}
\end{split} 
\end{align}
Note that Eq.~\ref{eq:doppler} follows from the assumption that $\frac{\partial x_2}{\partial y_2} = \frac{\partial z_1}{\partial y_1} = 0$. This assumption is equivalent to stating that the lensed images are stationary transverse to the lens. Alternatively, this is equivalent to stating that the lensed images are only weakly magnified, which for the $A_3$ lens follows from the fact that the lens map is essentially flat away from the folds (see Fig.~\ref{fig:lensmap}). This assumption fails only for a small region near the folds. We are also assuming that the two lens screens are stationary relative to each other. Namely, we assume that the velocity $\frac{\partial {\bm y}}{dt}$ is dominated by the velocity of the source relative to the combined position of the two screens. 

We now have the tools to simulate the conjugate wave-spectra of our $A_3$ lens plus scattering screen system. The top and middle panel of the right column of Fig.~\ref{fig:datasims} shows the location of the lensed images in Doppler-delay space. The plotted points correspond to where the power would be localized in the conjugate wave-field. The conjugate wave-field for the actual observation is shown in the left column as comparison. For our simulation, we have chosen the dimensionless parameters $\alpha = 0.7$, $\sigma = 0.05$, $\nu = 31,000$, $D = 5$, $y_1 = 0$, and $y_2 = 11$. The locations of the images on the scattering screen are chosen to be distributed uniformly random over a range $z_1 \in [-16, 16]$. These values are chosen to be consistent with the observing frequency $f = 311\,{\rm MHz}$ and the distances, $d_{01} = 389\,{\rm pc}$, $d_{02} = 415\,{\rm pc}$, and $d_{03} = 620\,{\rm pc}$ measured by \citet{2022arXiv220806884Z}. We also choose the velocity to be in the direction $\frac{\partial {\bm y}}{\partial t} \parallel [1, 0.1]$ to be consistent with the velocity measured by \citet{2022arXiv220806884Z}. In order to convert the dimensionless time delay and Doppler shift to dimensionful quantities, we take the physical scale of the lens to be $\ell = 1\,{\rm AU}$, and the magnitude of the relative velocity of the source to the lens to be $v = 23\,{\rm km \, s^{-1}}$ (again, consistent with the velocity measured by \citet{2022arXiv220806884Z}).  From the parameters, we can also infer the value for the amplitude of the plasma under-density, $\Sigma_2^* \sim 0.0001\,{\rm pc}\,{\rm cm}^{-3}$. It is important to note that these chosen parameters are not the best-fit parameters for this model, but rather a reasonable estimate of the parameters in order to qualitatively compare the features of the model with the data. 

There are two features of the data that we wish to point out that our model naturally reproduces. Firstly, we note that the millisecond feature is highly asymmetric. That is, if the two lenses responsible for the main scattering screen and the millisecond feature were truly one-dimensional (i.e. translationally invariant along one axis), then the millisecond feature would simply be a lensed copy of the main parabolic arc, centred at a different delay and Doppler shift. This, however, is not the case; the millisecond feature is truncated as it moves towards larger Doppler shifts. This truncation is naturally accommodated by our model as the $A_3$ lens, by design, ends in a cusp. The folds that produce the lensed images meet at the cusp rather than continuing on forever. Secondly, as the lensed images approach this truncation point (the cusp) in Doppler-delay space, the separation between pairs of images in delay first becomes larger before converging at the cusp. This is the characteristic hockey stick shape that can be seen both in the data and simulation in the middle row of Fig.~\ref{fig:datasims}. This increase in the delay between images as one approaches the cusp is a generic feature of our model as the contours of the lens map effectively form a loop around the cusp, as seen in Fig.~\ref{fig:contours}. That is the images have a tendency to spread out before converging at the cusp.

We may also wish to examine the behaviour of the magnification of the images. The bottom left panel of Fig.~\ref{fig:datasims} shows the total flux of the millisecond feature as a function of observing frequency, computed by summing the total power in the millisecond feature, normalized to a value of unity at some reference frequency. Essentially what happens is that as the frequency increases, pairs of localized islands of power in the conjugate wave-field merge successively. When each pair merges they rapidly increase in brightness before disappearing. This happens multiple times as one varies the frequency, creating the structure in seen in the bottom left panel of Fig.~\ref{fig:datasims}: each spike in magnification corresponds to one of these mergers.  

Although we have not undertaken the more involved process of quantitatively determining the best-fit parameters of our model to the data, we hope to have demonstrated that qualitatively the $A_3$ lens can naturally accommodate features of the PSR\,B0834+06 event that have previously been challenging to accommodate. 

\section{ESEs and Scintillation}
\label{sec:scint}

One of the more powerful aspects of the doubly catastrophic lensing framework is that it has the potential to explain both ESEs and scintillation with a single, unified framework. While in this work we have focused on cusp ($A_3$) lenses as a natural explanation for ESEs, previous works have explored the possibility of explaining scintillation observations with ensembles of fold ($A_2$) lenses \citep{2014MNRAS.442.3338P, simard_predicting_2018}. The basic picture we propose is that corrugated plasma sheets are responsible for both scattering phenomena. When corrugated sheets are closely aligned with the line of sight, they form folds under projection. These folds result in the multi-path propagation that is seen in pulsar scintillation. While these folds are required to be highly elongated (i.e. effectively one-dimensional) to explain scintillation observations, they cannot continue forever. When folds end, they are mathematically required to merge in cusp ($A_3$) catastrophes. It is these $A_3$ catastrophes that we propose as the origin of ESEs. 

One immediate question that arises is why, if both scintillation and ESEs are caused by the same ISM structures, is one phenomenon so much more common than the other. Within the PSR\,B0834+06 observation we have been discussing, there is only one feature associated with an $A_3$ lens (namely, the millisecond feature), whereas the each of the scattered images along the main scintillation arc, in our picture, would be associated with a fold. Since there are hundreds of scattered images, this suggests that fold lenses are much more common than cusp lenses. Moreover, most pulsars are observed to scintillate, but are only observed to undergo ESE-like scattering about one percent of the time. 

While this relative rarity of ESEs compared to scintillation might initially seem to pose an issue for any attempt to explain these two phenomena with a unified model, the doubly catastrophic framework actually provides a natural explanation. It is a well-known result of catastrophe theory that the cross-section for folds is much larger than the cross-section for cusps. That is, consider a projected plasma surface density profile given by $\Sigma_e$. We can define the area on the sky such that the density is greater than some threshold, $\Delta$, to be $\sigma(\Sigma_e > \Delta)$. The scaling as a function of threshold for this cross-section can be computed for the fold and cusp catastrophes as \citep{1993LIACo..31..217N}:
\begin{align}
    \sigma_{A_2}(\Sigma_e > \Delta) &\sim \Delta^{-2}, \\
    \sigma_{A_3}(\Sigma_e > \Delta) &\sim \Delta^{-5/2}.
\end{align}
That is, the cross-section for the cusp decreases faster as a function of threshold than the fold cross-section. Therefore, it is a generic expectation of catastrophe theory that folds will contribute more to the observed density than cusps. This is also, notably, a precise and testable prediction of our model; the number of ESEs with an inferred maximum column density above some threshold should scale according to this power law.

\section{Applications}
\label{sec:applications}

We have argued that doubly catastrophic lensing is a potentially powerful framework for analyzing scattering phenomena in pulsars and other radio sources, as it provides a unified explanation for both scintillation and ESE observations, and also naturally accommodates qualitative features of the data that have thus far been challenging to explain. Another powerful aspect of lenses as catastrophes is that the mathematics of catastrophe theory constrains the form of the lenses to a small set of elementary catastrophes. These elementary catastrophes are universal and are described by a small number of unfolding parameters. 

If the plasma structures in the ISM responsible for these scattering phenomena are indeed catastrophes, then this would represent a significant advancement in our ability to unambiguously infer the physical properties of the lenses and also to use them as astrometric tools. Contrast this with the present situation. Since we lack any prior information on the form of the lenses, in principle one has an infinite number of degrees-of-freedom when attempting to build a model to match ESE or scintillation data. As a result, inferences of, say, the electron density, $\Sigma_e$, may vary by orders-of-magnitude between different lens models. Moreover, there has been little observational evidence, so far, that has allowed us to distinguish between the many proposed models. At the very least, the doubly catastrophic lensing formalism makes precise predictions which we will be able to test soon. If confirmed, then the space of potential lens shapes collapses from infinite, to a small number of catastrophes.

This would have particularly important implications for pulsar astrometry. One of the primary limitations of our ability to use lensing to obtain precise astrometric data is the fact that we typically do not know the dispersive contribution of the lens to the observed time delays. That is, we are typically forced to ascribe the observed time delays entirely to the geometric part of the time delay, e.g. the quadratic terms in Eq.~\ref{eq:delay}. If the lenses are catastrophes, described by a small number of parameters, then it becomes possible to unambiguously infer the dispersive contributions to the delay. Especially for a system such as PSR\,B0834+06 which is highly over-determined, it would potentially be possible to infer the full lens potential.

\begin{figure}
    \centering
    \includegraphics[width=\columnwidth]{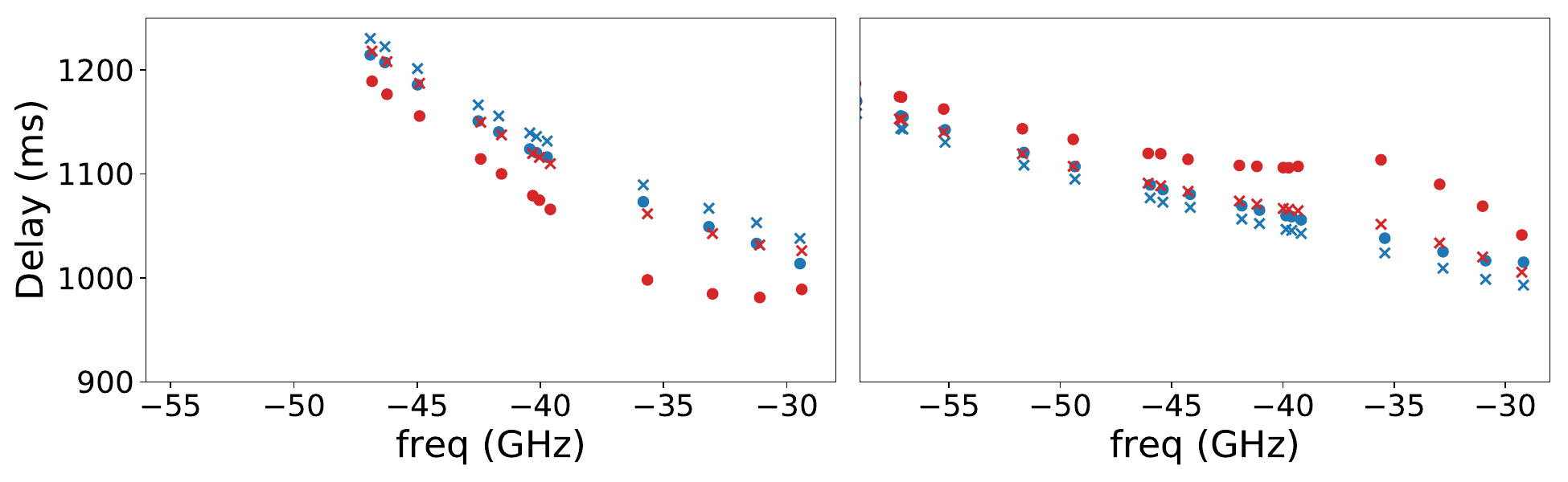}
    \caption{The simulated millisecond feature using the same parameters for our double-lens model as Fig.~\ref{fig:datasims}, except the left panel is for $\alpha > 0$ (the under-dense lens) and the right panel is for $\alpha < 0$ (the over-dense lens). The crosses show the value of the geometric delay, whereas the solid dots show the total delay (geometric plus group delay). The red and blue colours indicate the parity of the images, which is either positive or negative, given by the sign of the determinant of the Jacobian of the lens map.}
    \label{fig:div_vs_conv}
\end{figure}

One concrete example of an application of the doubly catastrophic lensing framework to infer the physics of scattering structures in the ISM is its potential ability to unambiguously distinguish between under-dense and over-dense lenses. Fig.~\ref{fig:div_vs_conv} shows the millisecond feature of the PSR B0834+06 lensing event, modelled with the $A_3$ lens for using $\alpha > 0$ (left panel) and $\alpha < 0$ (right panel). The crosses show the value of the geometric delay, whereas the solid dots show the total delay. Note that since the group delay is negative for positive $\alpha$, and positive for negative $\alpha$, for an under-dense lens, the total delay is less than the geometric delay, and for an over-dense lens, the total delay is greater than the geometric delay. The red and blue colours indicate the parity of the images, which refers to the direction of motion of the images with respect to the source. If the image moves in the same direction as the source then it is said to have positive parity, and if it moves in the opposite direction it is said to have negative parity. Note that in either the over- or under-dense case, the positive parity images have much smaller group delay than the negative parity images. As a result, the orientation of the characteristic hockey-stick shape of the millisecond feature is flipped between the over-dense and under-dense lenses. Inspecting the data shown in Fig.~\ref{fig:datasims} visually, the millisecond feature appears to be closer to the under-dense case. However, since we have not attempted to precisely fit the data, it is not possible to make any strong inferences. 

An additional possibility that the doubly catastrophic framework opens up, is that if pulsar scintillation is caused by $A_2$ folds and ESEs are caused by $A_3$ cusps by the same sheet structures in the ISM as viewed under projection, then many observations of these phenomena will allow us to take advantage of the rich mathematics of catastrophe theory to probe the turbulent ISM on scales that are inaccessible to simulations. That is, with many scintillation and ESE observations, we can effectively generate a sky-map of caustic networks in the ISM. Such a map may be a powerful tool for studying the physics of the ISM.

\section{Conclusion}
\label{sec:conclusion}

In this paper, we have presented a model based on a simple application of catastrophe theory to thin plasma sheets to explain extreme scattering events. That is, we propose that several aspects of ESE observations can be explained using lens potential with an $A_3$ cusp profile. This is an extension of previous work \citep{2014MNRAS.442.3338P, simard_predicting_2018} suggesting that $A_2$ folds arising from corrugated plasma sheets may explain pulsar scintillation. We call this application of catastrophe theory to the lens potential ``doubly catastrophic" lensing, as catastrophes also generically appear in the light curves of lensed sources. 

The doubly catastrophic framework is well-motivated for several reasons. Firstly, the past decade of pulsar scintillation observations suggest the ubiquity of thin plasma sheets in the ISM. Since lensing is well-described by an effective projected density perpendicular to the line of sight, and since catastrophes generically arise when thin sheets are viewed under projection, $A_2$ and $A_3$ lenses (in addition to higher order catastrophes which have not considered here) should naturally arise. We argue that these catastrophic lens potentials should exist in the ISM, whether or not they are abundant enough to explain all scintillation or ESE observations. Secondly, recent work on the physics of the turbulent ISM through MHD simulations suggest that corrugated plasma sheets of the kind we consider are physically well-motivated. Lastly, the doubly catastrophic framework has several desirable theoretical features: it provides a universal framework that describes both scintillation and ESEs as aspects of the same phenomenon, and the application of catastrophe theory means that the lens potentials are generic and well-described by a small number of parameters.

In this work, we have described the features of the simplest $A_3$ lens and have argued that it can explain many of the qualitative features of ESE observations, including the frequency structure of ESEs. The inability to account for features of ESE light curves at high frequencies has been a roadblock for thin sheet models of these events. We argue that the $A_3$ lens overcomes this issue. We also argue that the $A_3$ lens provides a natural explanation for features seen in the lensing of PSR B0834+06.

\section*{Data Availability}
No new data were generated or analysed in support of this research.

\section*{Acknowledgements}
We receive support from Ontario Research Fund—research Excellence Program (ORF-RE), Natural Sciences and Engineering Research Council of Canada (NSERC) [funding reference number RGPIN-2019-067, CRD 523638-18, 555585-20], Canadian Institute for Advanced Research (CIFAR), Canadian Foundation for Innovation (CFI), the National Science Foundation of China (Grants No. 11929301),  Thoth Technology Inc, Alexander von Humboldt Foundation, and the Ministry of Science and Technology(MOST) of Taiwan(110-2112-M-001-071-MY3). Computations were performed on the SOSCIP Consortium’s [Blue Gene/Q, Cloud Data Analytics, Agile and/or Large Memory System] computing platform(s). SOSCIP is funded by the Federal Economic Development Agency of Southern Ontario, the Province of Ontario, IBM Canada Ltd., Ontario Centres of Excellence, Mitacs and 15 Ontario academic member institutions.
 Cette recherche a \'{e}t\'{e} financ\'{e}e par le Conseil de recherches
en sciences naturelles et en g\'{e}nie du Canada (CRSNG), [num\'{e}ro de
r\'{e}f\'{e}rence 523638-18,555585-20 RGPIN-2019-067]. 

\bibliographystyle{mnras_sjf}
\bibliography{biblio} 



\appendix


\bsp	
\label{lastpage}
\end{document}